\documentstyle[twocolumn,floats,aps]{revtex}

\begin{document}
\draft

\twocolumn[\hsize\textwidth%
\columnwidth\hsize\csname@twocolumnfalse\endcsname

\title{\bf A Discrete Model for Nonequilibrium Growth
Under Surface Diffusion Bias}

\author{S. Das Sarma and P. Punyindu}
\address{Department of Physics, University of Maryland, College Park, 
MD 20742-4111}

\date{\today}
\maketitle

\begin{abstract}
A limited mobility nonequilibrium solid-on-solid dynamical model 
for kinetic surface growth is introduced as a simple description for the
morphological evolution of a growing interface under random vapor
deposition and surface diffusion bias conditions.
Simulations using a local coordination dependent
instantaneous relaxation of the deposited atoms produce complex surface
mound morphologies whose dynamical 
evolution is inconsistent with {\it all the
proposed continuum surface growth equations}. 
For any finite bias, mound
coarsening is found to be only {\it an initial transient}
which vanishes asymptotically, with the asymptotic
growth exponent being $0.5$ in both 1+1 and 2+1
dimensions.
Possible experimental implications of the proposed
limited mobility nonequilibrium model for real
interface growth under a surface diffusion bias
are critically discussed.
\end{abstract}

\vskip 1pc]
\narrowtext

An atom moving on a free surface is known to encounter an additional 
potential barrier, often called a surface diffusion bias \cite{1}, 
as it approaches
a step from the upper terrace --- there is no such extra barrier for an
atom approaching the step from the lower terrace (the surface step
separates the upper and the lower terrace). Since this diffusion bias
makes it preferentially more likely for an atom to attach itself to the
upper terrace than the lower one, it leads to mound (or pyramid) - type
structures on the surface under growth conditions as deposited atoms
are probabilistically less able to come down from upper to lower
terraces. This dynamical growth behavior 
is sometimes called an ``instability'' because 
a flat (``singular'') two dimensional 
surface growing under a surface diffusion bias is unstable
toward three dimensional mound/pyramid formation. 
There has been a great deal of recent interest 
\cite{1,4,5,6,7,7p,8,9,10,11,12,13,14,15,16,17p,17,18,19,20,22} in the 
morphological evolution of growing interfaces under nonequilibrium 
growth conditions in the presence of such 
a surface diffusion bias. 
In this paper we propose a {\it minimal}
nonequilibrium cellular automata - type atomistic growth model for 
ideal molecular beam epitaxial -
type random vapor deposition growth under a surface diffusion
bias. Extensive stochastic simulation results presented in this paper
establish the morphological evolution of a surface growing under 
diffusion bias conditions to be surprisingly complex
even for this extremely simple minimal model. 
Various critical growth exponents, which {\it asymptotically}
describe the large-scale dynamical evolution of the
growing surface in our minimal discrete growth model, are
inconsistent with {\it all} the proposed continuum theories for
nonequilibrium surface growth under diffusion bias conditions. Our
results based on our extensive study of this minimal model 
lead to the conclusion that a continuum description for
nonequilibrium growth under a surface diffusion bias does not
exist (even for this extremely simple minimal model)
and may require a theoretical formulation which is substantially
different from the ones currently existing in the literature.
Our results in the initial non-asymptotic transient growth
regime (lasting upto several hundred or a few thousand
layers of growth) do, however, agree with existing
theoretical and (many, but not all) 
experimental findings in the literature.

In Fig. 1(a) we schematically show our solid-on-solid (SOS) 
nonequilibrium growth model : (1) Atoms are deposited randomly (with
an average rate of 1 layer/unit time, which defines the unit of time in
the growth problem --- the length unit is the lattice spacing taken to be
the same along the substrate plane and the growth direction) and
sequentially on the surface starting with a flat substrate; (2) a
deposited atom is incorporated instantaneously if it has at least one
lateral nearest-neighbor atom; 
(3) singly coordinated deposited atoms (i.e. the ones
without any lateral neighbors) could instantaneously relax to a
neighboring site within a diffusion length of $l$ provided the
neighboring site of incorporation has a higher coordination than the
original deposition site; (4) the instantaneous relaxation process is
constrained by two probabilities $P_L$ and $P_U$ ($0 \leq P_L,P_U
\leq 1$) where $P_{L(U)}$ is the probability for the atom to attach itself
to the lower(upper) terrace after relaxation (note that a ``terrace'' here
could be just one other atom). The surface diffusion bias is implemented
in our model by taking $P_U > P_L$, making 
it more likely for atoms to attach to the upper terrace.
Under the surface diffusion bias, therefore, 
an atom deposited at the top of a
step edge feels a barrier (whose strength is controlled by $P_U/P_L$) in
coming down compared with an atom at the lower terrace attaching itself
to the step. Our model is 
well-defined for any value of the
diffusion length $l$ including the most commonly studied situation of
nearest-neighbor relaxation ($l=1$).   
(We should emphasize, however, that the definition of 
a surface diffusion bias is not unique even within our
extremely simple limited mobility nonequilibrium growth
model --- what we study in this Letter is the so-called
edge diffusion bias \cite{12}.)
\begin{figure}

 \vbox to 4.5cm {\vss\hbox to 6cm
 {\hss\
   {\includegraphics{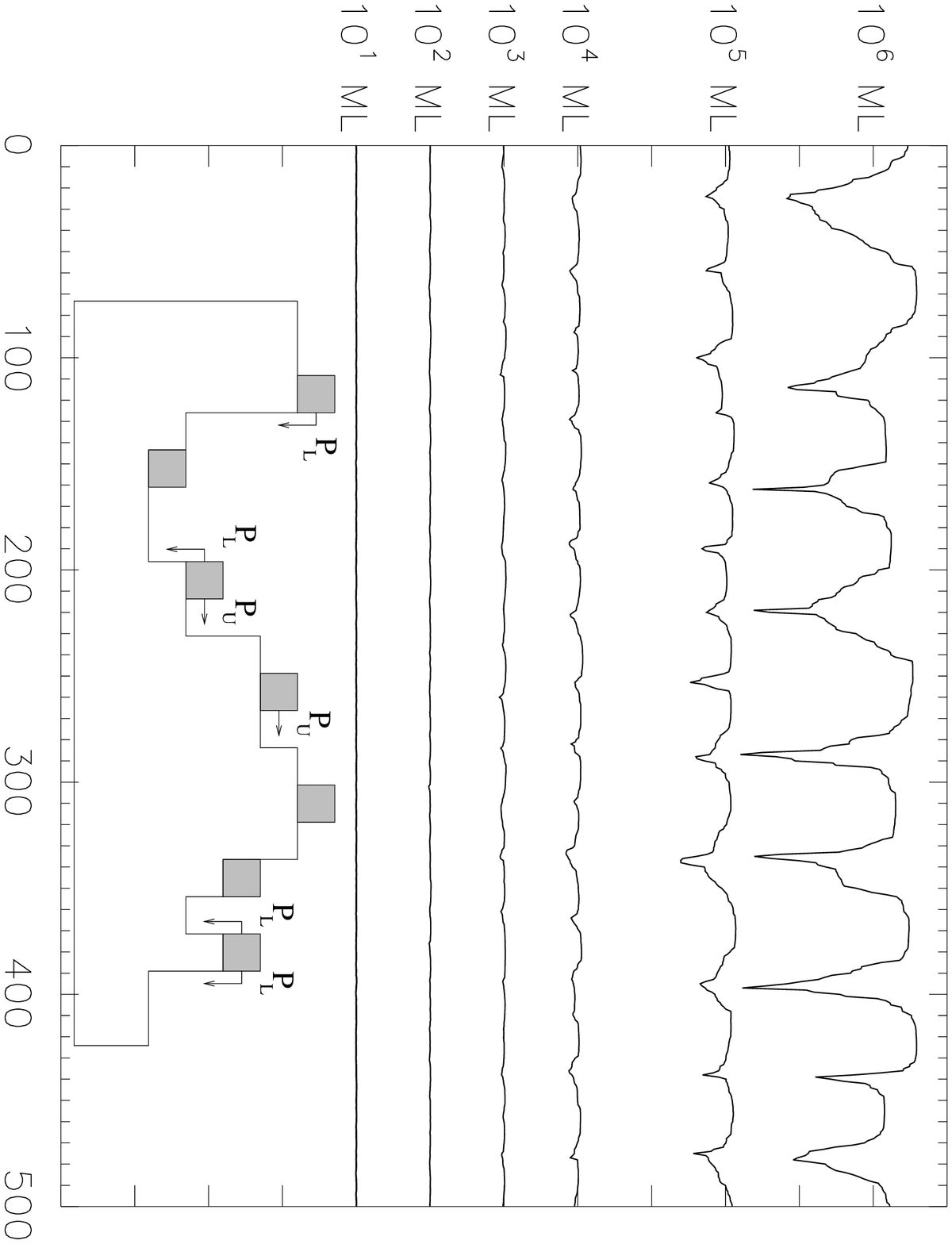}
   }
  \hss}
 }

 \vbox to 4.5cm {\vss\hbox to 6cm
 {\hss\
   {\includegraphics{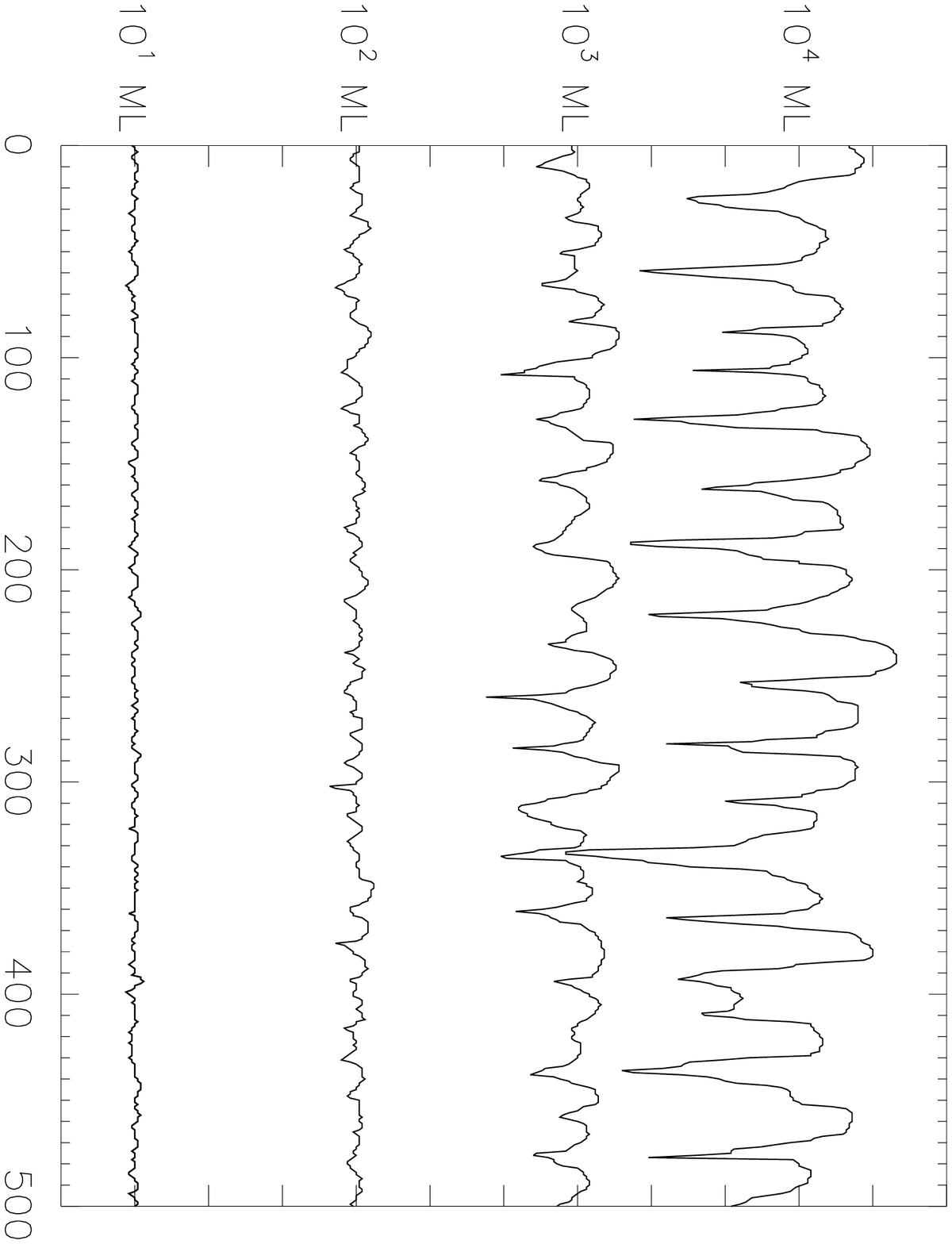}
   }
  \hss}
 }

 \vbox to 4.5cm {\vss\hbox to 6cm
 {\hss\
   {\includegraphics{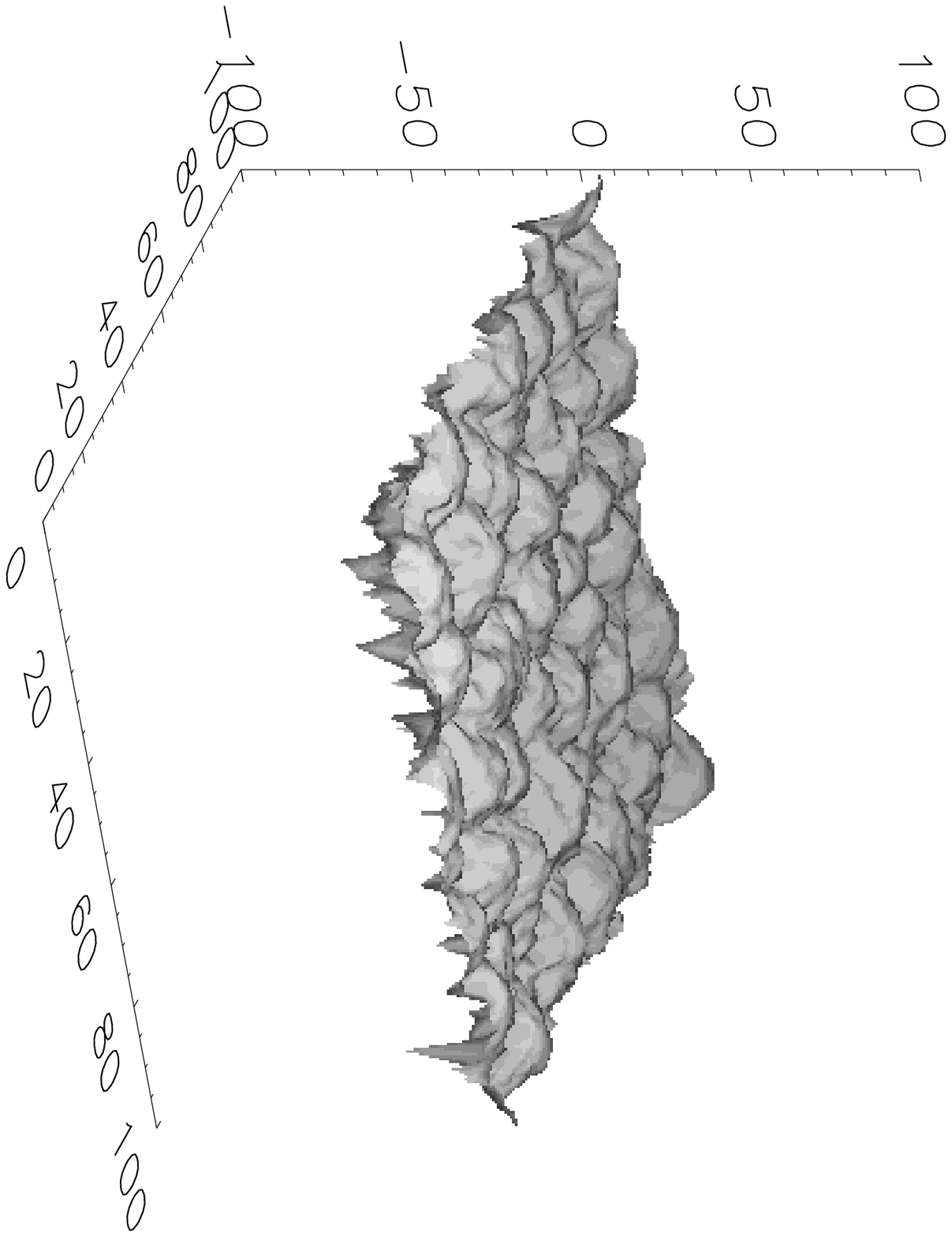}
   }
  \hss}
 }

 \vbox to 4.5cm {\vss\hbox to 6cm
 {\hss\
   {\includegraphics{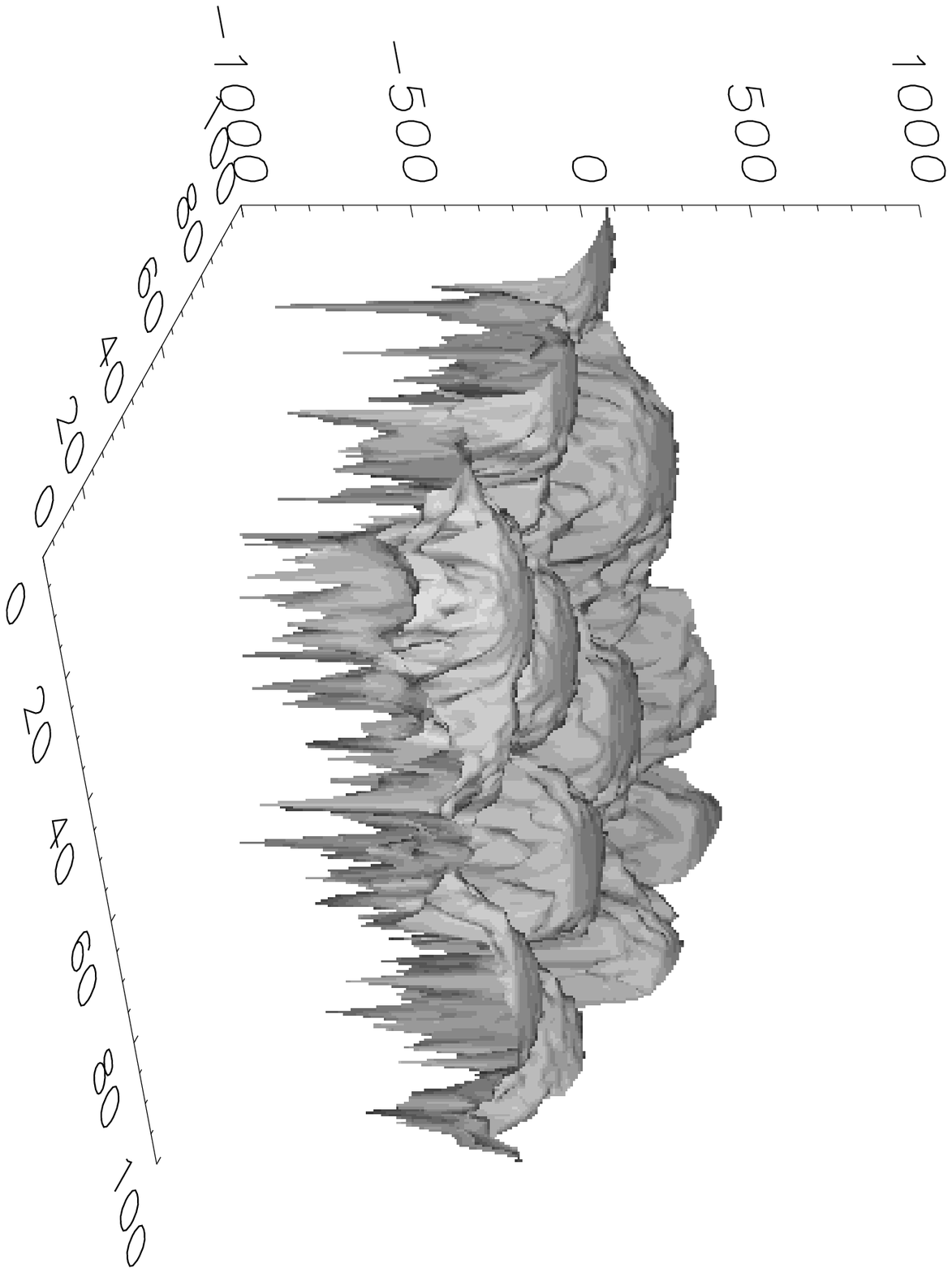}
   }
  \hss}
 }

\caption{
(a) The mound evolution in 1+1 dimensions
showing a section of $500$ middle lattice sites from a
substrate size of $L=10000$ at $10-10^6$ ML
($P_U=1;P_L=0.5$).
Lower inset: Schematic configuration defining growth rules
in 1+1 dimensions.
(b) The details of the initial mound evolution in (a)
for $10-10^4$ ML, showing that the mounds form very early
in the growth. Somewhat noisy mounds can already be seen
at 10 ML, and the coarsening of the mounds between
$10-10^4$ ML is clearly seen in the results.
(c,d) The d=2+1 growth morphologies on a $100 \times 100$ substrate
($P_U=1;P_L=0.5$) at
(c) $10^3$ ML, and
(d) $10^6$ ML.
}
\end{figure}
We have carried out extensive simulations
both in $1+1$ and $2+1$ dimensions (d) varying $P_L$, $P_U$ as well as
$l$, also including in our simulations the inverse situation (the
so-called `negative' bias condition) with $P_L > P_U$ so that deposited
atoms preferentially come down attaching themselves to lower steps
producing in the process a smooth growth morphology.
Because of lack of space we do not present 
here our `negative' bias ($1 \geq P_L > P_U$) results
(to be published elsewhere)
except to note that the smooth dynamical growth morphology
under our negative bias model obeys 
exactly the expected \cite{1} linear Edwards - Wilkinson
universality \cite{22}.
Our growth model is the most obvious {\it finite bias} generalization
of the well-studied cellular automaton model referred to as the
DT model or the 1+ model in the literature \cite{22,23,24}.

Before presenting our numerical results we point out two important
features of our growth model : (1) For $P_L = P_U = 1$ our model reduces
to the one introduced in ref. 22, often called the DT model,  
(and studied extensively 
\cite{22,23,24,25,26,27}
in the literature) as a minimal model for molecular beam epitaxy 
{\it in the absence of any diffusion bias};
(2) we find, in complete
agreement with earlier findings \cite{23,24} in the absence of diffusion
bias, that the diffusion length $l$ is an {\it irrelevant} variable
(even in the presence of bias)
which does not affect any of our calculated critical exponents
(but does affect finite size corrections --- increasing $l$ requires a
concomitant increase in the system size to reduce finite size effects).
To demonstrate this, we compare two systems (see Fig. 2(c)), 
one with diffusion length $l=1$ while the other with $l=5$.
Both systems are on sufficiently large substrates of size
$L=10^4$ to prevent finite size effect and both have the
same bias strength ($P_U=1.0$ and $P_L=0.9$).
Except for the expected layer-by-layer growth seen in the first few
layers in the $l=5$ system, we see the same critical
behavior from the two systems with the coarsening processes
{\it slightly} faster in the $l=5$ system in early time regime.
In the rest of this paper
(except for Fig. 2(c) where we present some 
representative $l=5$ results),
we present our $l=1$ simulation results 
emphasizing that our critical exponents are independent of $l$ provided
finite size effects are appropriately accounted for. Our calculated 
exponents are also independent of the precise values of $P_U$ and
$P_L$ ( $< P_U$ )
as found \cite{23,24} in the unbiased $P_U=P_L$ case.

In Fig.1 we show our representative d=1+1 (a and b)
and 2+1 (c and d) simulated dynamical growth
morphology evolution. The diffusion 
bias produces mounded structures which are visually statistically scale
invariant only on length scales much larger 
(or smaller) than the typical mound size.
Note that the mounding in the growth morphology starts
very early during growth and is already prominent in the
first 100 monolayers (ML) as is obvious in Fig. 1(b).
In producing our final  
results we utilize a noise
reduction \cite{28} technique which accepts only a fraction of the
attempted kinetic events, and in the process produces smoother results
(reducing noise effects)
without affecting the critical exponents.
We have explicitly checked that the noise reduction technique
does not change our calculated critical exponents \cite{28}
and has only the cosmetic effect of suppressing noise in our
simulated growth morphology.

To proceed quantitatively we now introduce the dynamic scaling ansatz
\cite{22,23,24}
which seems to describe well all our simulated results. 
\begin{figure}

 \vbox to 5.cm {\vss\hbox to 6cm
 {\hss\
   {\includegraphics{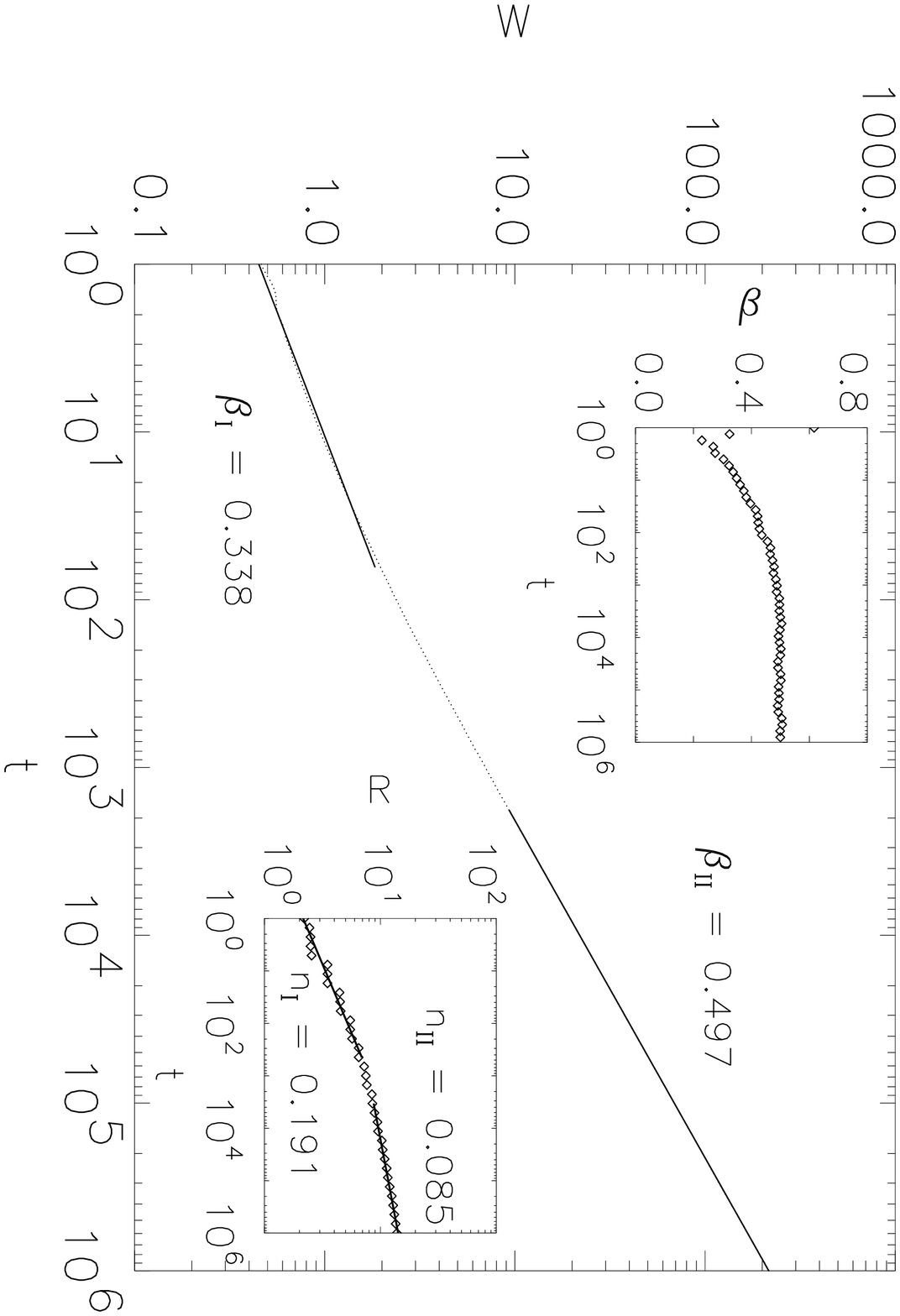}
   }
  \hss}
 }

 \vbox to 5.cm {\vss\hbox to 6cm
 {\hss\
   {\includegraphics{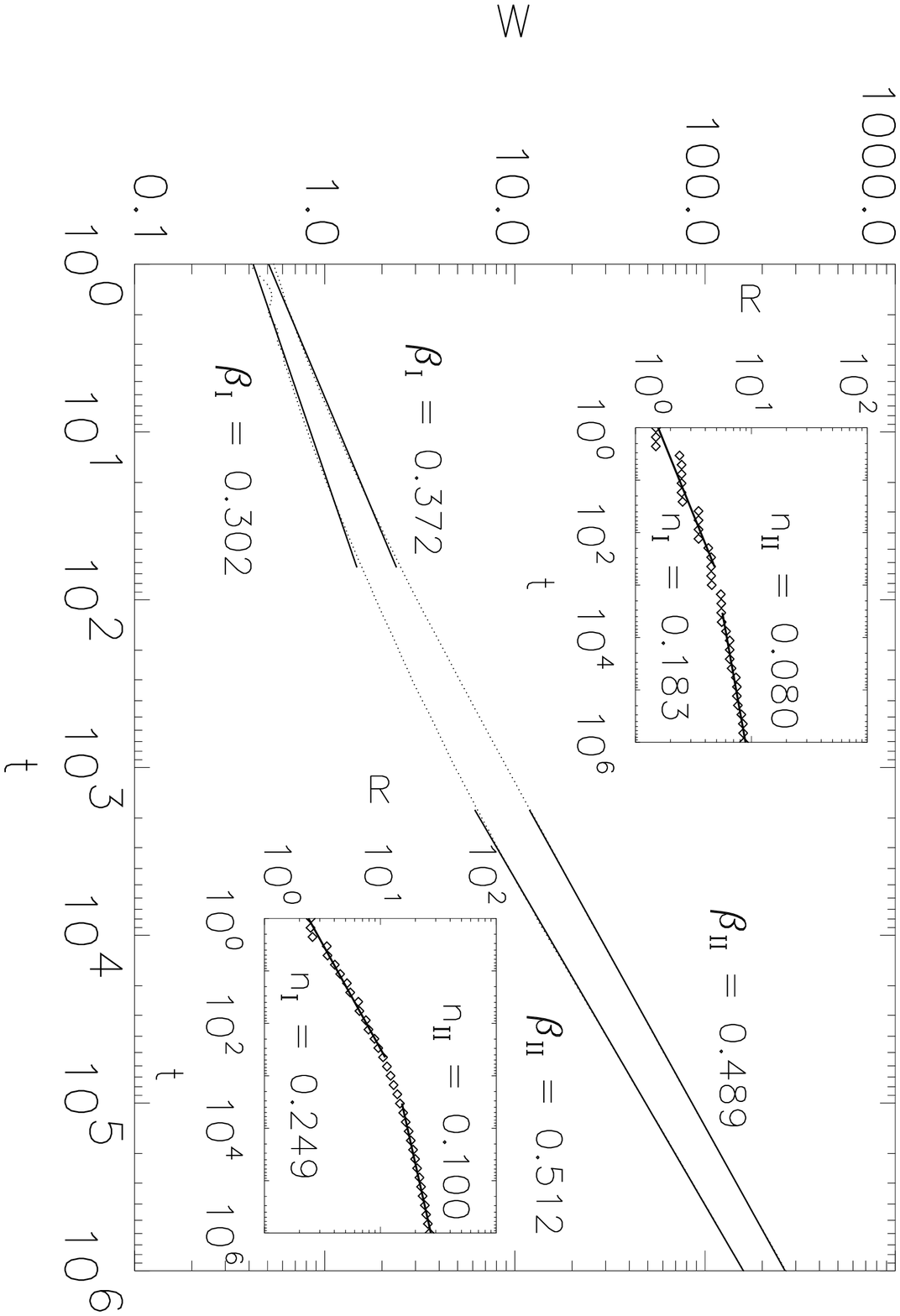}
   }
  \hss}
 }

 \vbox to 5.cm {\vss\hbox to 6cm
 {\hss\
   {\includegraphics{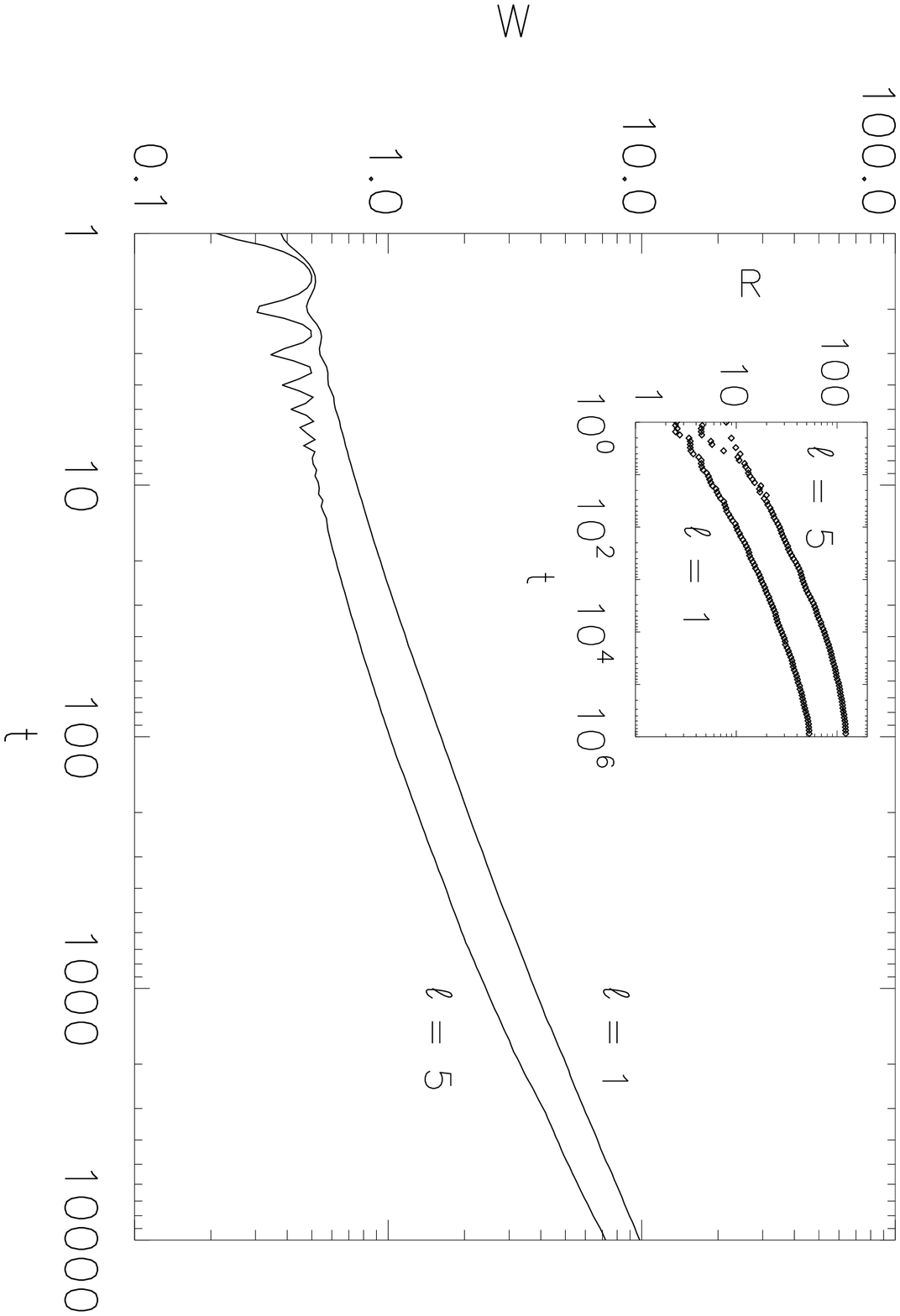}
   }
  \hss}
 }

\caption{
(a) The surface roughness $W$
in the $L=10000$ system with $P_U=1;P_L=0.5$.
Left inset: the growth exponent $\beta$
calculated from the local derivative of $\log_{10}W$ with respect to
$\log_{10}t$. Right inset: Average mound size vs
time.
(b) $W$ in
$L=10000$ systems with $P_U=1;P_L=0.25$ (upper curve) and
$P_U=1;P_L=0.75$ (lower curve).
Left inset: Average mound size vs time for
the system with $P_U=1;P_L=0.25$.
Right inset: The same plot for $P_U=1;P_L=0.75$.
(c) Results for diffusion length $l=1$ compare with $l=5$
(all other results in this paper correspond to $l=1$).
The results for $l=1$ and $l=5$ are parallel to each other
for both $W$ and $R$ (inset) as a function of time,
indicating exponent universality with respect to the
diffusion length. In this figure, $P_U=1.0; P_L=0.9$.
The bending in the $W(t)$, $R(t)$ curves in $\log-\log$
plots shown here indicates the crossover behavior from the
initial transient regime to the asymptotic regime as
discussed in the text and shown in Fig. 2 (a) and (b).
The layer-by-layer oscillation can clearly be seen
upto 10 (3) ML for $l=5$ ($l=1$).
}
\end{figure}
We have studied
the root mean square surface width or surface roughness ($W$), the
average mound size ($R$), the average mound height ($H$), and the
average mound slope ($M$) as functions of growth time. We have also
studied the various moments of dynamical height-height correlation
function, and these correlation function results (to be reported elsewhere)
are consistent with the ones obtained from our study of $W(t)$, $R(t)$,
$H(t)$, and $M(t)$. The dynamical scaling ansatz in the context of the
evolving mound morphologies can be written as power laws in growth time
(which is equivalent to power laws in the average film thickness) : $W(t)
\sim t^\beta$ ; $R(t) \sim t^n$ ; $H(t) \sim t^\kappa$ ; $M(t) \sim
t^\lambda$ ; $\xi(t) \sim t^{1/z}$, where $\xi(t)$ is the lateral correlation
length (with $z$ as the dynamical exponent) and $\beta$, $n$, $\kappa$,
$\lambda$, $z$ are various growth exponents which are not necessarily
independent. We  
find in all our simulations $n \simeq z^{-1}$, and thus the coarsening
exponent $n$, which describes how the individual mound sizes increase in
time, is the same as the inverse dynamical exponent in our model. We
also find $\beta = \kappa$ in all our results, which is understandable
in a mound-dominated morphology. In addition, all our results satisfy
the expected exponent identity $\beta = \kappa = n + \lambda$ because
the mound slope $M \sim H/R$. The evolving growth morphology is thus
completely defined by two independent critical exponents $\beta$ (the
growth exponent) and $n$ (the coarsening exponent), which is similar to
the standard (i.e. without any diffusion bias) dynamic scaling situation
\cite{22,23,24}
where $\beta$ and $z$ ($= n^{-1}$ in the presence of diffusion bias)
completely define the scaling properties. 

In Figs.2 and 3 we show our representative scaling results 
in d=1+1 (Fig.2) and 2+1 (Fig.3)
for nonequilibrium
growth under surface diffusion bias conditions. It is clear that we
consistently find the growth exponent 
$\beta \simeq 0.5$ in both d=1+1 and 2+1  
{\it in the long time asymptotic limit} independent of $P_L$ 
and $P_U$ as long as $P_L/P_U < 1$.
This $\beta \simeq 0.5$ is, however, different from the usual
Poisson growth under pure random deposition with no relaxation
where there are no lateral correlations.
Our calculated {\it asymptotic} coarsening 
exponent $n$ in both d=1+1 and
2+1 is essentially zero ($< 0.1$) at long times.
In all our results we find the effective 
coarsening (growth) exponent showing a
crossover from $n$ ($\beta$) $\approx 0.2$ ($0.25$ or
$0.33$ depending on d=2+1 or 1+1)
at early times ($1 < t \alt 10^3$)
to a rather small (large) value
( $n < 0.1$, $\beta \approx 0.5$ as $t \rightarrow \infty$)
at long times --- we believe the asymptotic $n$ 
($\beta$) to be zero (half) in our model.
Our calculated steepening exponent $\lambda$ satisfies
the exponent identity $\lambda = \beta - n$ rather well,
indicating that steepening ($M \sim t^{\lambda}$) and
coarsening ($R \sim t^{n}$) are competing processes.
We find that during the initial transient regime
(for $1 < t \alt 10^3$ depending on d and $P_L/P_U$),
when considerable mound coarsening takes place
($n \sim 0.2$), $\lambda$ is rather small 
($\lambda < 0.1$) and does not change much.
We note that the mound formation dominates our
growth morphology even during the initial 
transient --- mounding starts early, coarsens rapidly,
and then coarsening slows down or almost stops.
After the initial transient, however, $\lambda$
is finite and large ($\lambda \agt 0.4$), indicating
significant steepening of the mounds. Based on
our simulation results we are compelled to conclude
that the evolutionary behavior of our mound dynamics
makes a crossover (at $t=t^{\ast} \alt 10^3$) from 
a coarsening-dominated preasymptotic regime $I$ ($n_I \sim 0.2$,
$\lambda_I < 0.1$) to a steepening-dominated regime ${II}$
($n_{II} < 0.1$, $\lambda_{II} \agt 0.4$) which we
believe to be the asymptotic regime.
The initial transient (regime $I$: $t<t^{\ast}$) can be
construed as a ``slope selection'' regime where the
steepening exponent $\lambda$ is very small (and the
coarsening exponent, $n \sim 0.2$, 
is approximately a constant), but the asymptotic
long time behavior (regime ${II}$: $t>t^{\ast}$) is
clearly dominated by slope steepening (large $\lambda$)
with coarsening essentially dying down ($n < 0.1$).
The crossover time $t^{\ast}$ between regime $I$ 
(coarsening) and $II$ (steepening) depends on the details 
of the model (e.g. d, L, $P_L$, $P_U$, $l$), and could
be quite large ($t^{\ast} \sim 10^2 - 10^4$). 

\begin{figure}

 \vbox to 5.cm {\vss\hbox to 6cm
 {\hss\
   {\includegraphics{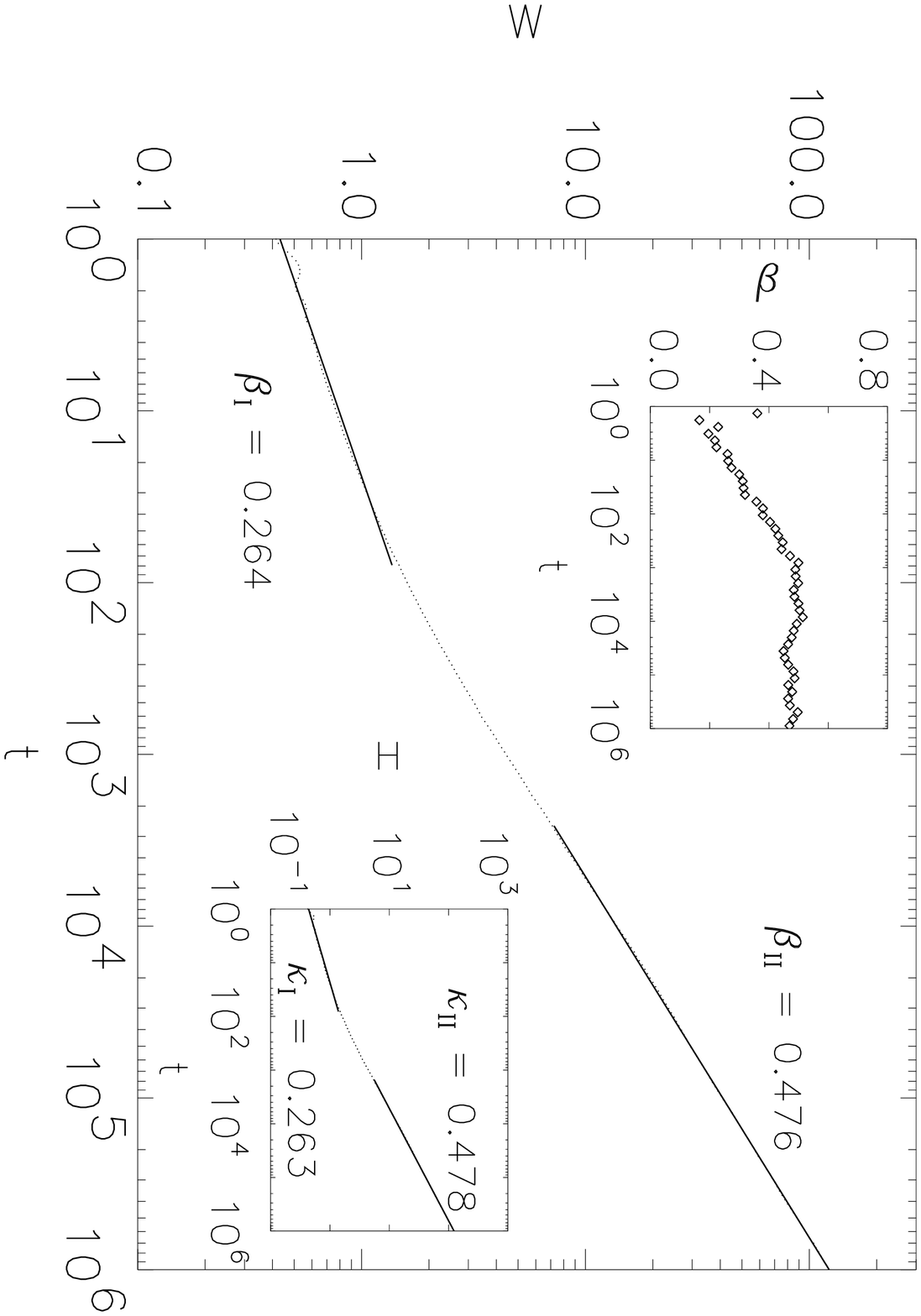}
   }
  \hss}
 }

 \vbox to 5.cm {\vss\hbox to 6cm
 {\hss\
   {\includegraphics{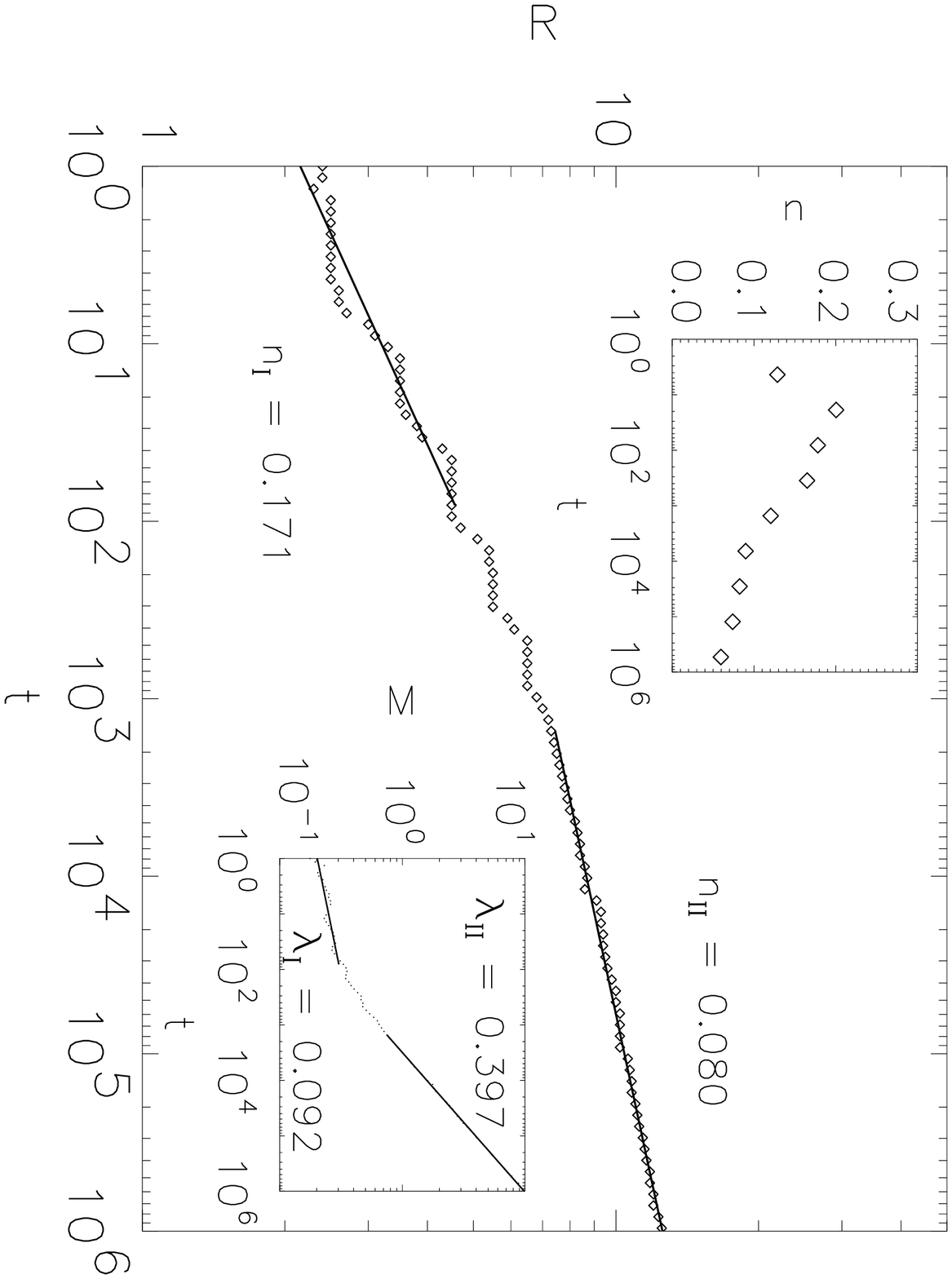}
   }
  \hss}
 }

\caption{
(a) The surface roughness $W$ in the $100 \times 100$ system
with $P_U=1;P_L=0.5$. Left inset:
the local growth exponent $\beta$. Right inset: Average mound height vs
time.
(b) The average mound size in the
same $100 \times 100$ system. Left inset:
the local coarsening exponent $n$ calculated from the local derivative
of $\log_{10}R$ with respect to $\log_{10}t$. Right inset: the average
mound slope vs time.
}
\end{figure}
We emphasize that our initial transient
($t<t^{\ast}$) exponents in regime $I$ ($\beta_I \simeq 0.26$;
$n_I \simeq 0.17$; $\lambda_I \simeq 0.09$ in d=2+1,
Fig.3) agree {\it quantitatively} with several \cite{7,7p,16}
non-minimal detailed (temperature-dependent Arrhenius
diffusion) growth simulations as well as with direct numerical 
simulations \cite{11} of a proposed (empirical) continuum
growth equation, which have all been claimed
\cite{7,7p,11,16} to be in good agreement with the 
observed coarsening behavior in various epitaxial growth
experiments. 
Other simulations and theories \cite{6,17}
find no slope selection, which also agree with some experiments.
We believe that the observed slope selection ($\lambda \approx 0$) in
experiments and simulations is a (long-lasting) transient
(our regime $I$, $t<t^{\ast}$) behavior, which should disappear 
in the asymptotic regime,
at least in models without any artificial
transient downward mobility.
Theoretically, a distinction has been made
\cite{17p,17,19,22} between `weak' and `strong' 
diffusion bias cases, corresponding respectively to our 
regime $I$ (`mounding') and $II$ (`steepening'), respectively.
Our results show that this theoretical distinction is not
meaningful because the `weak' bias case crosses over to
the `strong' bias case for $t>t^{\ast}$, and the long time
asymptotic regime is invariably the `strong' bias
steepening regime for any finite diffusion bias.
(There could be accidental slope selections at
rather large slopes when crystallographic orientations
are taken into account \cite{13,22}, a process neglected in 
our minimal growth model.)

In comparing with the existing {\it continuum growth equation} results 
we find that none can quantitatively
explain all our findings. 
Golubovic \cite{18} predicts $\beta = 0.5$, which is consistent with our
asymptotic result ($\beta_{II} \simeq 0.5$),
but his finding of $n = \lambda = 1/4$ in both d=1+1 and
2+1 is inconsistent with our asymptotic results 
(asymptotically $n < 0.1$, and $\lambda \simeq 0.4-0.5$)
while being approximately consistent with our initial
transient results (for $t<t^{\ast}$). The analytic results of
Rost and Krug \cite{19} also cannot explain our results, because they
predict, in agreement with Golubovic, that if $\beta = 1/2$, then $n =
\lambda = 1/4$.  
We also find our asymptotic 
$\beta$ to be essentially $0.5$ independent of the
actual value of $P_U/P_L$, which disagrees with ref. 19.
Interestingly our {\it initial transient regime} is, in fact,
approximately consistent with the theory of ref. 19.
We have approximate
slope selection ($\lambda < 0.1$)
only in the initial transient regime, which could, however,
be of considerable experimental relevance because the
pre-asymptotic regime is a long-lasting transient.
The only prior work in the literature that has some similarity to
our results is that of Villain and collaborators \cite{6,10},
who in a {\it one dimensional deterministic} (i.e. {\it without}
the deposition beam shot noise) macroscopic continuum
description of growth roughness and instabilities 
under surface diffusion bias,
called the Zeno model \cite{6,10} by 
the authors, found, in agreement with our atomistic 
{\it two dimensional stochastic} cellular automaton 
simulation results, 
a scenario in which coarsening 
becomes ``extremely slow after the mounds have reached a
(characteristic) radius'' \cite{10}, which is reminiscent of our
crossover from weak bias ($t<t^{\star}$) to strong bias
($t>t^{\star}$). While the precise relationship between
our two (and one) dimensional atomistic/stochastic model
and their \cite{6,10} one dimensional macroscopic/deterministic
model is unclear at the present time, it is interesting 
to note that the authors of refs. \cite{6,10} came to 
a similar negative conclusion as we do about the non-existence
of any continuum growth equation describing the strong bias
asymptotic regime where the microscopic lattice size may
play a crucial role \cite{26}. Finally, we note that a
very recent experimental work \cite{R} reports growth 
morphology of Ge(001) surface which agrees qualitatively
with the scenario predicted in this paper, namely, that
even for a very weak diffusion bias, the mound slope
continues to increase without any observable coarsening.

Although our diffusion bias model is an extremely simple
limited mobility model which may be viewed as unrealistic,
our d=1+1 results are remarkably similar to those obtained from a study
of a full temperature-dependent Arrhenius hopping model 
with step edge barrier \cite{12}.
This study \cite{12} offered essentially the same
picture as what we present here, i.e. a smaller value of
the growth exponent that crosses over to $\beta \approx 0.5$
at larger time, and a very large dynamical exponent
corresponding to very little coarsening 
($n \rightarrow 0$) after approximately
100 monolayers of deposition.
To be specific, the dynamical exponent determined
by the growth of the correlation length is 
$z \approx 16.6$ in the temperature-dependent edge bias
model \cite{12} while our study yields $z \agt 10$.
Although the dynamical exponents from the two studies
are not {\it exactly} the same, they are both
exceptionally large indicating the mound coarsening
process to be negligible in the large time regime.
This qualitative agreement between our simple
limited mobility results and a d=1+1 full diffusion results
argue strongly in favor of our minimal growth model
being of reasonably general qualitative validity
in experimental situations.

Finally, we note that the introduction of 
limited mobility models \cite{23} 
has opened a whole new
way of studying kinetic surface roughening in molecular
beam epitaxial growth. These limited mobility nonequilibrium models 
make it possible to study very large systems in very large 
time limit when it is impossible to do so in the realistic
temperature-dependent full-activated diffusion models.
Particularly, we point out the success of the DT model
\cite{23} in providing an excellent zeroth order description of molecular
beam epitaxial growth in the absence of any surface 
diffusion bias. The d=2+1 critical exponents \cite{27}
in the unbiased (DT) model \cite{23},
which belongs to the same universality class as the conserved
fourth-order nonlinear continuum MBE growth equation \cite{27,28},
are $\beta = 0.25 - 0.2$ and $\alpha \simeq 0.6-0.7$, which are in
quantitative agreement with a number of experimental measurements
\cite{22} where surface diffusion bias is thought to be
dynamically unimportant.
With this in mind, 
it is, therefore, conceivable that our study of 
the limited mobility model
which includes surface diffusion bias
(i.e. a generalized version of the DT \cite{23} model)
presented in this paper may benefit the subject
in a way similar to what the DT model did for the unbiased
molecular beam epitaxy growth study. 
This is particularly significant since there
are still many open questions 
regarding the interface growth under surface diffusion
bias condition.  
Since the model we study here is a generalized DT model \cite{23} 
and an approximate
continuum description for the original unbiased model \cite{23}
has recently been developed
\cite{26}, one could use that as the starting point to construct a
continuum growth model for the biased growth situation.
Such a continuum description is, however, extremely complex \cite{26} as
it requires the existence of an infinite number of
nonlinear terms in the growth
equation, and it therefore remains unclear whether a meaningful
continuum description for our discrete simulation results
is indeed possible \cite{6,10,note}.

In conclusion, we want to emphasize the fact that the 
limited mobility model studied in this paper is an extremely
simple model 
(``{\it the minimal model}'' in the sense that it is perhaps the
simplest nonequilibrium model which captures the minimal
features of growth under surface diffusion bias),
and realistic growth under experimental conditions should be
substantially more complex than this minimal model
(we speculate that this minimal model is in the same growth
{\it universality class} as realistic growth under a surface
diffusion bias for reasons discussed above, but we certainly
cannot prove that at this incomplete stage of the development
of the subject.
A word of caution is in order in comparing experimental
results with our calculated critical exponents
because of the extreme simplicity and the limited
mobility nature of our nonequilibrium growth model.)
If we do not have a reasonable theoretical understanding
(from a continuum equation approach) of even such a
simple minimal model, as we have found in this paper,
then current efforts 
\cite{1,4,5,6,7,7p,8,9,10,11,12,13,14,15,16,17p,17,18,19,20,22}
at understanding realistic growth under surface diffusion bias
must be quite futile.
The main weakness of limited mobility model
(of the type presented here)
is that they are manifestly nonequilibrium model --- this,
however, should not be a particularly serious problem in
the context of the mound/pyramid formation in the growth
morphology, which by definition is a nonequilibrium effect
and must disappear in a properly equilibrated surface.
Our conclusion based on the results presented in this
Letter is that a continuum growth equation for nonequilibrium
growth under a surface diffusion bias does not exist
at the present time.

This work is supported by the US-ONR and the NSF-DMR-MRSEC.


\begin{thebibliography}{33}
\bibitem{note} Our speculation about the difficulty
of constructing a continuum description for nonequilibrium growth
under surface diffusion bias conditions is in fact consistent
with the earlier speculation by Krug, p.240 in ref. 21, who
concluded that the `strong' bias case ``cannot be expected to be 
well-described by continuum theory'' --- our all important new
finding is that any finite diffusion bias, however weak, 
invariably renormalizes to the `strong' bias situation with
asymptotic $\beta \simeq 0.5$ and $\lambda \simeq 0$, and all
earlier `weak' bias coarsening behavior findings in the 
literature are only pre-asymptotic ($t<t^{\ast}$) transients.
\end{thebibliography}
\end{document}